\documentclass[aps,pre,twocolumn,floatfix,superscriptaddress,showpacs,showkeys]{revtex4-1}
\usepackage{amsmath,amssymb,verbatim,color,multirow,pifont}
\usepackage{graphicx}
\usepackage{braket,commath}
\usepackage{hyperref}
\usepackage{cleveref}
\usepackage{enumerate}
\usepackage[utf8]{inputenc}

\newcommand{\Order}[1]{O{(#1)}}
\newcommand{\Net}{\mathcal{L}}
\newcommand{\Forest}{\mathcal{F}}
\newcommand{\Spanning}{\mathcal{S}}
\newcommand{\Active}{\mathcal{A}}
\newcommand{\Inactive}{\mathcal{I}}
\newcommand{\Added}{\mathcal{D}}

\newcommand{\Edges}{\mathcal{E}}

\newcommand{\KMean}{k}

\crefname{equation}{Eq.}{Eqs.}
\crefname{section}{Sec.}{Secs.}
\crefname{figure}{Fig.}{Figs.}
\crefname{table}{TABLE}{TABLES}
\crefname{appendix}{Appendix}{Appendices}

\makeatletter
\AtBeginDocument{\@ifpackageloaded{natbib}{\ifNAT@numbers\if@filesw\immediate\write\@auxout{\string\global\string\NAT@numberstrue}\fi\fi}{}}
\makeatother
\begin{document}

\setcounter{page}{1}
\title[]{An efficient algorithm to compute mutually connected components in interdependent networks}
\author{S. Hwang}
\affiliation{Institute for Theoretical Physics, University of Cologne, 50937 K\"oln, Germany}
\affiliation{CCSS and CTP, Department of Physics and Astronomy, Seoul National University, Seoul 151-747, Korea}
\author{S.~\surname{Choi}}
\author{Deokjae~\surname{Lee}}
\author{B. Kahng}
\affiliation{CCSS and CTP, Department of Physics and Astronomy, Seoul National University, 
Seoul 151-747, Korea}
\email{bkahng@snu.ac.kr}
\date[]{Received \today}

\begin{abstract}
Mutually connected components (MCCs) play an important role as a measure of resilience in the study of interdependent networks. Despite their importance, an efficient algorithm to obtain 
the statistics of all MCCs during the removal of links has  thus far been absent. Here, using a well-known fully dynamic graph algorithm, we propose an efficient algorithm to accomplish this task. 
We show that the time complexity of this algorithm is approximately $\Order{N^{1.2} }$ for random graphs, which is more efficient than $\Order{N^{2}}$ of the brute-force algorithm. 
We confirm the correctness of our
algorithm by comparing the behavior of the order parameter as links are removed with existing
results for three types of double-layer multiplex networks. 
We anticipate that this algorithm will be used for simulations of large-size systems that have been previously inaccessible.  
\end{abstract}

\pacs{89.75.Hc, 64.60.ah,05.10.-a}

\maketitle

{\it Introduction.}$-$ Networks are ubiquitous in our world, and many of these interact with one another ~\cite{Barabasi1999,Barrat2008,Pastor-Satorras2007,Dorogovtsev2013}.
One striking instance of a strong internetwork correlation was the power outage in 2003 in Italy \cite{Buldyrev2010}, 
during which the power grid network and a computer network strongly interacted with each other.
A failure in one network thus led to another failure in the other network and this process continued back and forth. 
Such avalanche processes can continue until no additional node can fail. 
This avalanche of failures and its devastating consequences triggered efforts
to assess the resilience of interdependent network structures against external forces
\cite{Buldyrev2010,Parshani2010,Parshani2011,Gao2011,Hu2011,Huang2011,Li2012,Gao2012,Zhou2013,Son2011}.
        
As a natural measure of resilience of such interdependent networks, the size of a mutually connected component (MCC) per system size has  served as an order parameter of the percolation transition  ~\cite{Buldyrev2010,Son2012,Bianconi2014,Min2014,Watanabe2014}.
Here the MCC means that a node belonging to an MCC is connected to all other nodes directly or 
indirectly in the same MCC in   each layer network, called the A-layer network and the B-layer network, respectively.
Note that each node in the A-layer network has a one-to-one correspondence to its counterpart node in B-layer network. 
However, each layer network has its own set of link connections between nodes, and these are  
independent of those of the other layer network. Although MCCs have been proven to be an excellent 
measure of network resilience, obtaining results for large-size systems 
 has been computationally difficult because of the absence of an efficient algorithm. This problem was partially solved by a recently 
proposed algorithm in which a newly designed data structure was used to keep track of the size of a giant MCC 
efficiently during removal processes of nodes \cite{Schneider2013}. However, one still needs to resort to the brute-force algorithm when other physical quantities such as the size distribution of the MCCs are requested.  
        
Here, we introduce another efficient algorithm that keeps track of not only the size of a giant MCC but also the 
sizes of all other MCCs, and thus the size distribution of MCCs can be traced during removal processes. Particularly, our algorithm is designed to proceed as links are deleted. Thus, the percolation transition of the 
size of a giant MCC can be traced in terms of the actual number of removed nodes. To design it, we utilized a fully dynamic graph algorithm widely used in the computer science community, the one introduced in \cite{Holm2001}, which is called the HDT algorithm hereafter.

For each layer network, 
	once a component that is not a tree is changed into a spanning tree,
	its connection profile is saved in a special type of data structure called an Euler tour (ET) tree \cite{Henzinger1995,Henzinger1999}, because ET trees can be efficiently managed to merge or 
split spanning trees. However, to maintain the spanning trees efficiently when link deletions occur, information of redundant paths between nodes needs to be organized properly. The HDT algorithm 
is a way to maintain such information. It guarantees amortized $\Order{\log^2 N}$ time for a link deletion or creation 
when the ET tree is used for the data structure of the spanning trees.
The details of the ET tree are presented in Appendix A.

\begin{figure*}
\includegraphics[width=0.95\linewidth]{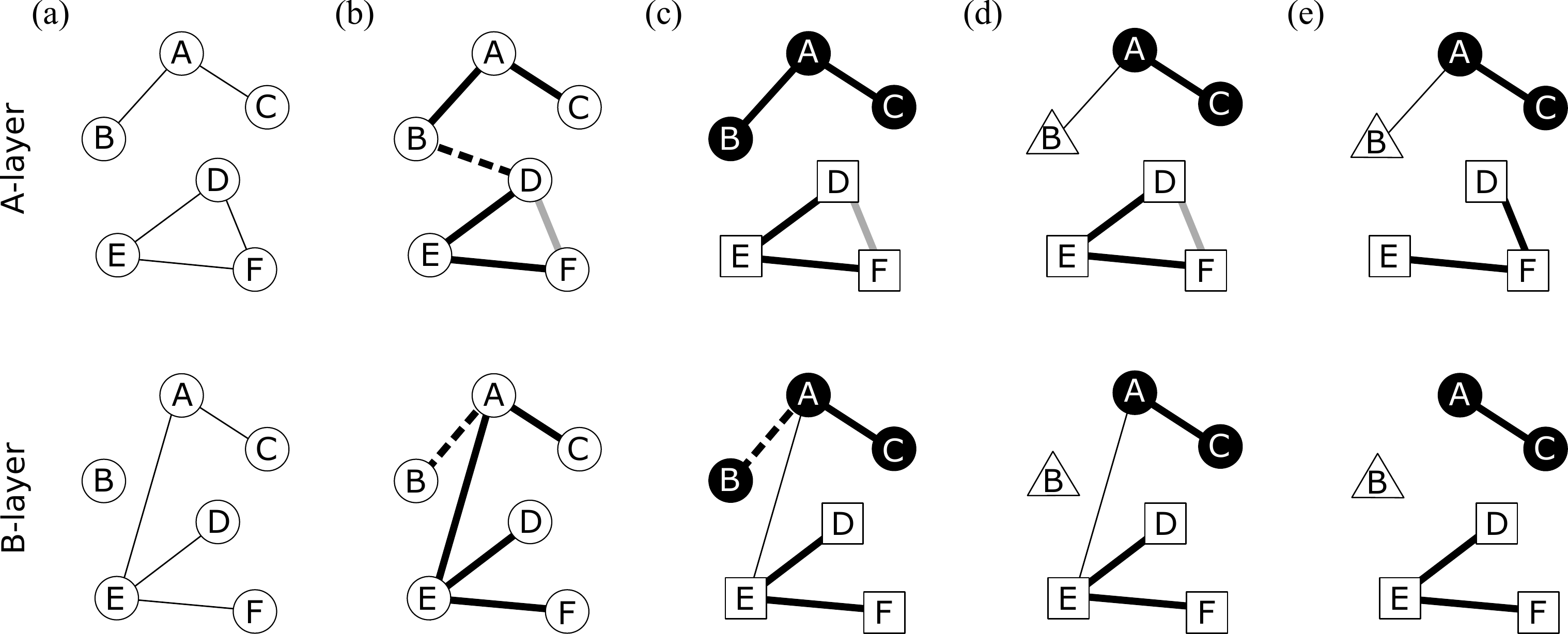}
\caption{
(a) Initial configurations of A-layer (upper) and B-layer (lower) networks. 
(b) First, each connected component is maintained in a spanning tree form. Link D--F (gray line) in the A layer 
is treated as a redundant link. 
Second, {\it ad hoc} links (dashed lines) B--D in the A layer and A--B in the B layer are added between 
two nodes through randomly selection from each component to connect the networks.
Then, there is only one MCC and all links including the {\it ad hoc} links are active (thick lines).
(c) An {\it ad hoc} link B--D is deleted in the A layer. This deletion splits the A-network into two components.
Subsequently,  link A--E in the B layer becomes inactive (thin line) and we identify two MCCs \{A, B, C\} and \{D, E, F\}.
(d) The other {\it ad hoc} link A--B in the B layer is deleted.
Subsequently,  link A--B in the A layer becomes inactive (thin line) and the component \{A, B, C\} is split into two components \{A, C\} and \{B\}.
At this stage, there are no  remaining {\it ad hoc} links and the MCCs (represented by different node symbols) of the networks in (a) have been retained with identification of active and inactive links. 
(e) Now we delete links in the original networks one by one in the same manner.
Here we show two examples of link deletion that do not cause a cascade of inactivations:
(i) link E--D deletion in the A layer and (ii) link A--E deletion in the B layer.  For case (i) the redundant link D--F is recovered and maintains the spanning tree.
For case (ii), because the link is inactive, nothing occurs.
}
\label{Fig:diagram}
\end{figure*}

{\it Algorithm.}$-$ 
We first briefly introduce the prerequisites needed to explain our dynamic graph algorithm.
To query and update the connection profile of networks, we maintain a data structure for each layer in the form 
of ET trees. Those ET trees constitute a dynamic forest (DF) denoted as $\Forest$,
which performs the following four operations efficiently:
\begin{enumerate}
\item Connected($v$, $w$, $\Forest$) determines whether or not the nodes $v$ and $w$ are in the same component.
\item Size($v$, $\Forest$) returns the size $n$ of the component (tree) that contains the node $v$.
\item Insert($e$, $\Forest$) adds a link $e$ to $\Forest$.
\item Delete($e$, $\Forest$) removes a link $e$ from $\Forest$.
\end{enumerate}
All of these operations can be performed within $\Order{\log^2 N}$ computing time using the HDT algorithm.

Our algorithm consists of two main parts: identifying MCCs of a given multiplex network  
 and evolving the MCCs as links are removed one by one. 
Each update utilizes the previous information of the details of MCCs.
Throughout these processes, we trace the evolution of MCCs as a function of the number of links removed. 

To be specific, each link is categorized as active or inactive. Here, an active link is the one that belongs to an MCC. 
The rest of the occupied links are regarded as inactive. For example, the thick solid and dashed lines in \cref{Fig:diagram} represent active links, whereas the thin lines represent inactive ones. 
We remark that, even if two nodes $(v,w)$ are connected by a link $e$ in layer A and through a certain  
pathway in layer B, link $e$ can be inactive when the pathway in layer B contains one or more inactive links. 
However, once a link is deemed to be inactive, it remains inactive permanently throughout link removal processes. 
Using this simple fact, we design the algorithm to identify MCCs. 

In a double-layer multiplex network with $N$ nodes on each layer, let $\Net_A$ and $\Net_B$ denote the sets of links present on layers A and B, respectively. The DFs in each network are denoted as $\Forest_A$ and $\Forest_B$, respectively.
Each $\Forest_X$ (where $X$ represents either $A$ or $B$) stores the structure of MCCs of layer $X$ containing connection information of active links.

\begin{figure}[t]
\includegraphics[width=0.8\linewidth]{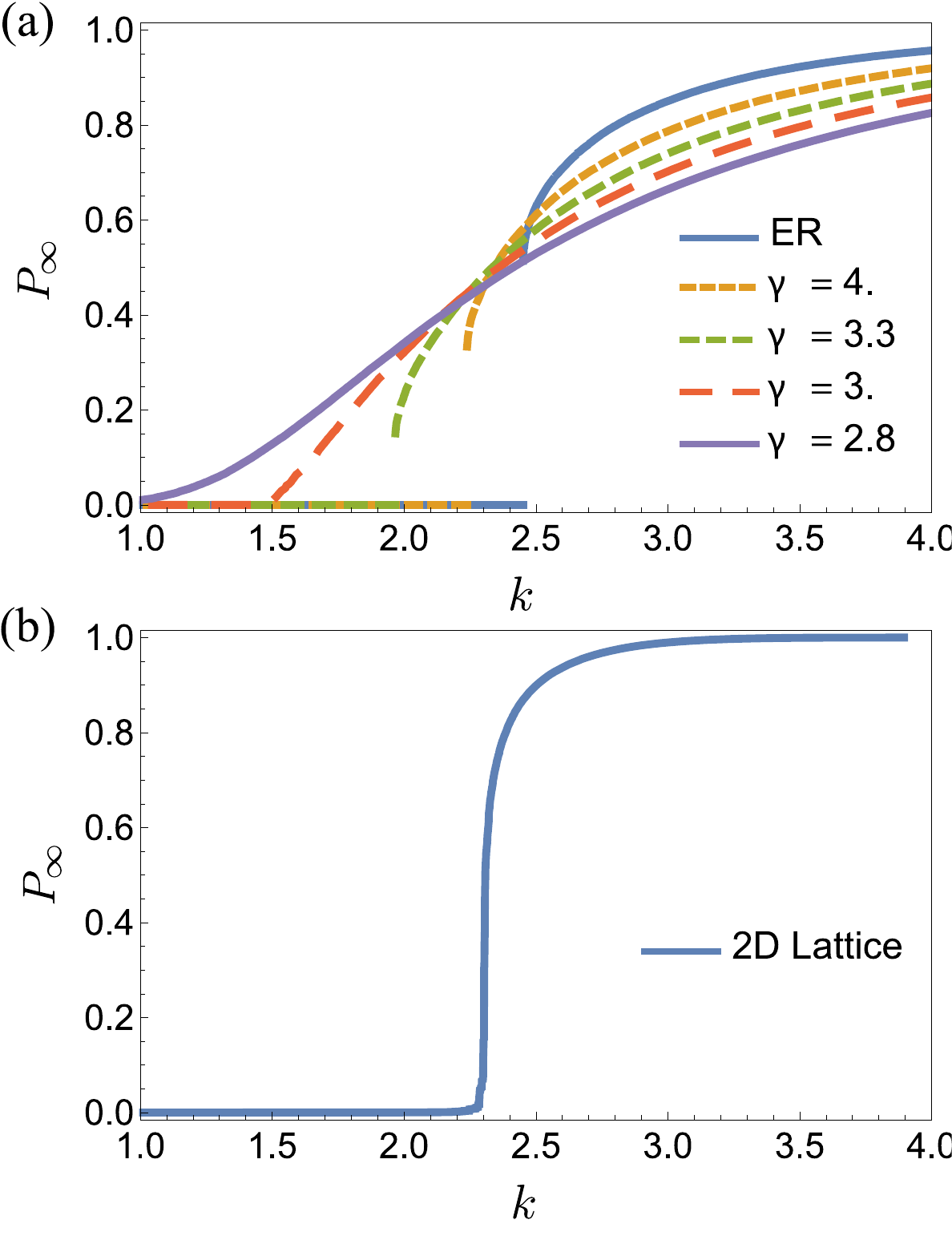}
\caption{(Color online)
Plot of $P_\infty$ (the size of a giant MCC divided by $N$) vs the mean degree $k=2L/N$, 
where $L$ is the number of remaining links in the system. $N=10^6$ 
and an initial mean degree $\KMean_0=4$ are taken. As links are removed randomly 
one by one from each layer, $P_\infty$ exhibits various discontinuous or continuous transitions depending on the underlying networks.
}
\label{Fig:OrderParam}
\end{figure}

\begin{figure}[t]
\includegraphics[width=0.8\linewidth]{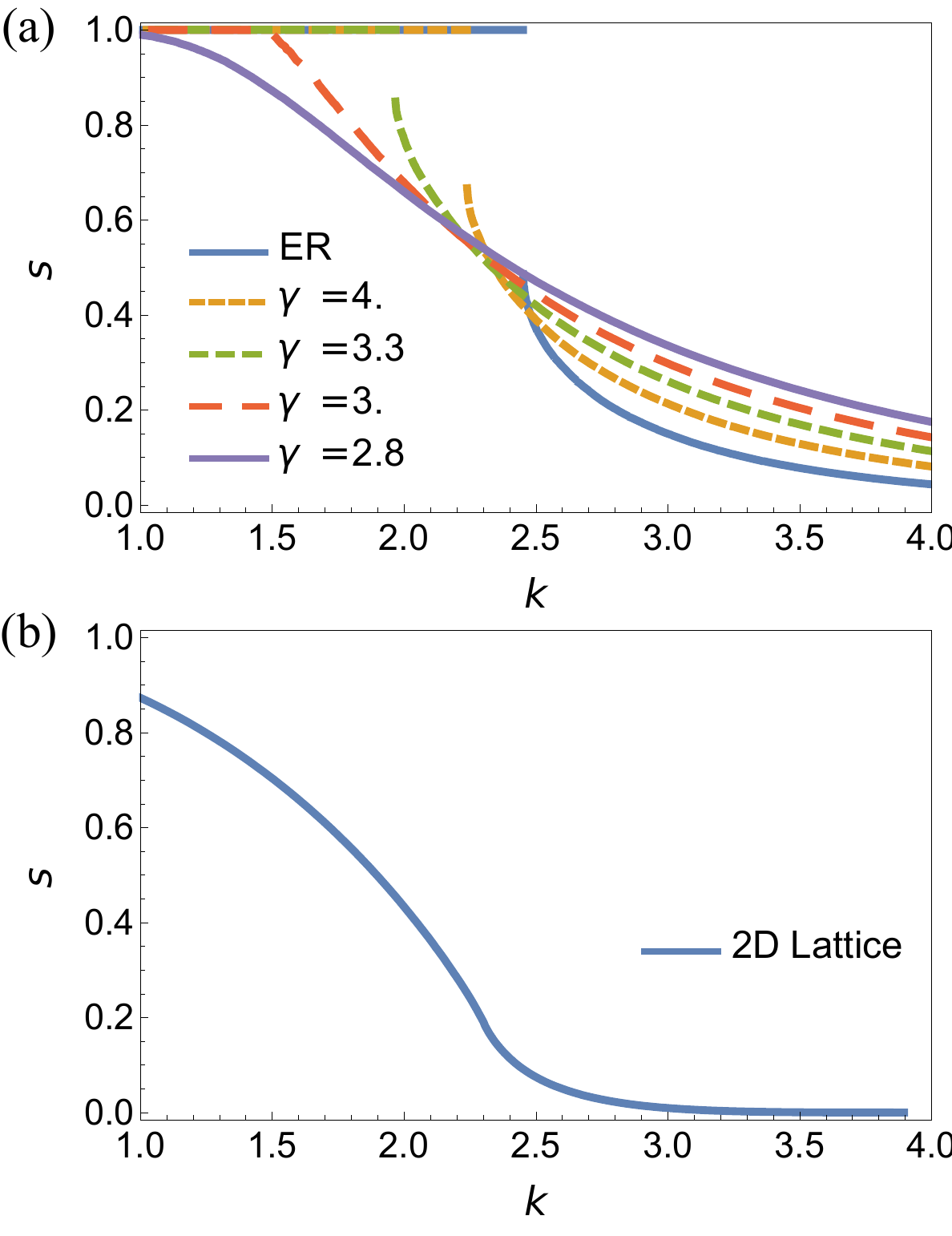}
\caption{(Color online) 
(a) Plot of $s$ (the number of MCCs divided by the system size $N$) vs 
the mean degree $k$ under the same conditions as those in Fig. \ref{Fig:OrderParam}. $s$ exhibits behavior similar to $m$ but in an upside-down manner for complex networks. 
(b) However, $s$ exhibits a somewhat different behavior from $m$ for the two-dimensional lattices.}
\label{Fig:NumCluster}
\end{figure}

The first part of the algorithm proceeds as follows: 
\begin{enumerate}[(i)]
\item For a given initial configuration of each layer network (see Fig. 1(a)), a spanning tree 
is extracted randomly from each component, based on which $\Forest_X$ ($X = \Set{A,B}$) is constructed. 
By using Connect($v, w, \Forest$), 
	the connection profile of each network is obtained.

\item 
To identify MCCs, some {\it ad hoc} links are added between disconnected trees, 
which means adding {\it ad hoc} links to $\Forest_X$.\footnote{
The standard ways are visiting each node of a network in a depth-first order or a breadth-first order.
Connected components are identified naturally in the visiting and {\it ad hoc} links between them can be created easily.
One simple way may be to pick up a node, from which links are added to the nodes that are not connected to it.
However, such a way is inefficient for simulation because a large number of {\it ad hoc} links are needed.
}
Let $\Added_X$ denote the set of all {\it ad hoc} links in layer $X$. 
Then, the set of active links can be denoted as $\Active_X=\Net_X \cup \Added_X$, and the set of inactive links becomes $\Inactive_X=\emptyset$ (Fig. 1(b)).

\item Choose a link at random from the set of {\it ad hoc} links $\Added_A$ and remove it (Fig. 1(c)).
If $e\in\Inactive_A$, then $e$ is removed from $\Inactive_A$.
This case does not occur at the beginning, but it can occur during the iterative processes.
If $e\notin\Inactive_A$, execute Delete($e = (v,w), \Forest_A$). 
If no other pathway connecting $v$ and $w$ exists, this component would be split into two.
The connection between $v$ and $w$ can be checked via Connect($v,w,\Forest_A$) after deleting $e$.

\item As shown in \cref{Fig:diagram}(d), the above division process in one layer may trigger some active links in the other layer into becoming inactive.
For each of those inactivated links $e$, execute Delete($e,\Forest_X$) and add it to $\Inactive_X$.
This deletion, in turn, can  trigger some other links in $\Active_Y$ into becoming inactive; then we repeat the above processes iteratively, where $Y$ represents the counterpart layer of $X$. The details can be found in \cref{Appendix:Removal}. 

\item Repeat  steps (iii) and (iv) until $\Added_A=\emptyset$ and $\Added_B = \emptyset$.

\end{enumerate}

After the above first part is completed, all MCCs for a given multiplex network are identified
and their structural information such as the sizes of each MCC can also be obtained from 
$\Forest_X (X=\{A,B\})$.
Next, we take the following step to determine the evolution of the MCCs as links are actually removed. 
This second part of the algorithm can be accomplished by taking  steps similar to (iii) and (iv).

\begin{enumerate}[(i)]
\setcounter{enumi}{5}
\item Repeat  steps (iii) and (iv) on $\Net_A$ and $\Net_B$ instead of $\Added_A$ and $\Added_B$. This process is repeated until the number of removed links reaches the value one wants (Fig. 1(e)). 
The order of link removals depends on the problem given. 
\end{enumerate}

Step (vi) contains the process of removals of active links in the original networks. Thus, we 
can trace the control parameter. Each updating of the second part builds on the 
DFs that store the MCC structures obtained in the previous iteration.
Therefore, by performing step (vi), we can easily garner the MCCs as a function of 
the number of remaining links. 

{\it Assessment.}$-$ We check the correctness of our algorithm for three types of double-layer multiplex networks: 
(i) random graphs proposed by Erd\H{o}s and R\'enyi (ER) \cite{Erdoes1959},
(ii) scale-free random graphs introduced in \cite{Goh2001},  in
which degree of a node in one layer is statistically the
same as the one of the corresponding node in the other
layer. Thus, degrees of nodes with the same node in-
dex on each layer are assortatively correlated \cite{Buldyrev2011}. 
In addition, we also consider (iii) two-dimensional regular lattices.
For each type of  network, the size of a giant MCC and the number of MCCs are measured as a function of 
the mean degree. Here the mean degree is given as $k=2L/N$, where $L$ is the number of links remaining in 
either $\Net_A$ or $\Net_B$ at each iteration step. Actually, those numbers are 
the same. To compare the results in a consistent manner, all networks are initiated by mean degree $k = 4$.

We first examine the size $P_{\infty}$ of a giant MCC normalized by the system size $N$  in \cref{Fig:OrderParam}(a).
For ER graphs, the order parameter exhibits a jump of $P_{\infty}^{\textrm{jump}} \approx 0.51$ at $\KMean_c \approx 2.46$, 
values of which are in agreement with the result in \cite{Buldyrev2010,Son2012}.
For scale-free networks, the jump sizes diminish as the degree exponent $\gamma$ decreases to 3.
For $\gamma \le 3$, the transition becomes continuous and no jump is obtained.
This result is also consistent with the previous result in \cite{Buldyrev2011} obtained using the conventional algorithm even though nodes are deleted there.

We also perform similar simulations for two-dimensional double-layer regular lattices. 
We obtain the transition point $k_c \approx 2.29$ in \cref{Fig:OrderParam}(b), corresponding to 
the occupation probability $p_c \approx 0.57$ in the conventional scheme.
This transition point is between $p_c = 0.5$ for the bond percolation transition and $p_c \approx 0.593$ for  
the site percolation transition in a two-dimensional monolayer network. Through the obtained numerical 
results thus far, we have confirmed that our algorithm successfully reproduces the previous results 
using the conventional algorithms.

We also examine the number $N_C$ of MCCs divided by the system size, $s = N_C/N$. One may expect that this quantity $s$ is 
small when a giant MCC exists in a large mean degree region; however, it is large when most of the links are deleted in a small mean degree region. 
The behavior of $s$ is shown in \cref{Fig:NumCluster}. 
For ER and scale-free networks, we find that $s$ exhibits a  behavior similar to that of $P_\infty$ in \cref{Fig:OrderParam} but 
in an upside-down manner. However, for the two-dimensional case, it looks somewhat different. 
The examination of the behavior of $s$ vs $k$ in such large-size systems would not be possible unless 
our algorithm is applied.

Next we consider the time complexity of the algorithm.  Step (v) forces steps (iii) and (iv) to be 
repeated $\Order{N}$ times, and for each (iii) and (iv), the Delete operation has to be executed at least once.
These steps request the computing time $\Order{N\log^2 N}$. However, when one link is deleted, 
inactivation of a multiple number of links can follow, which may induce cascading of deletions of links 
back and forth between the two layers. However, it is also possible that the deletion of one link may 
not induce any further inactivation process. Such complicated processes make it difficult to extract time complexity in 
a closed form. Thus, we resort to numerical methods to estimate time complexity. 

\begin{figure}
\includegraphics[width=0.95\linewidth]{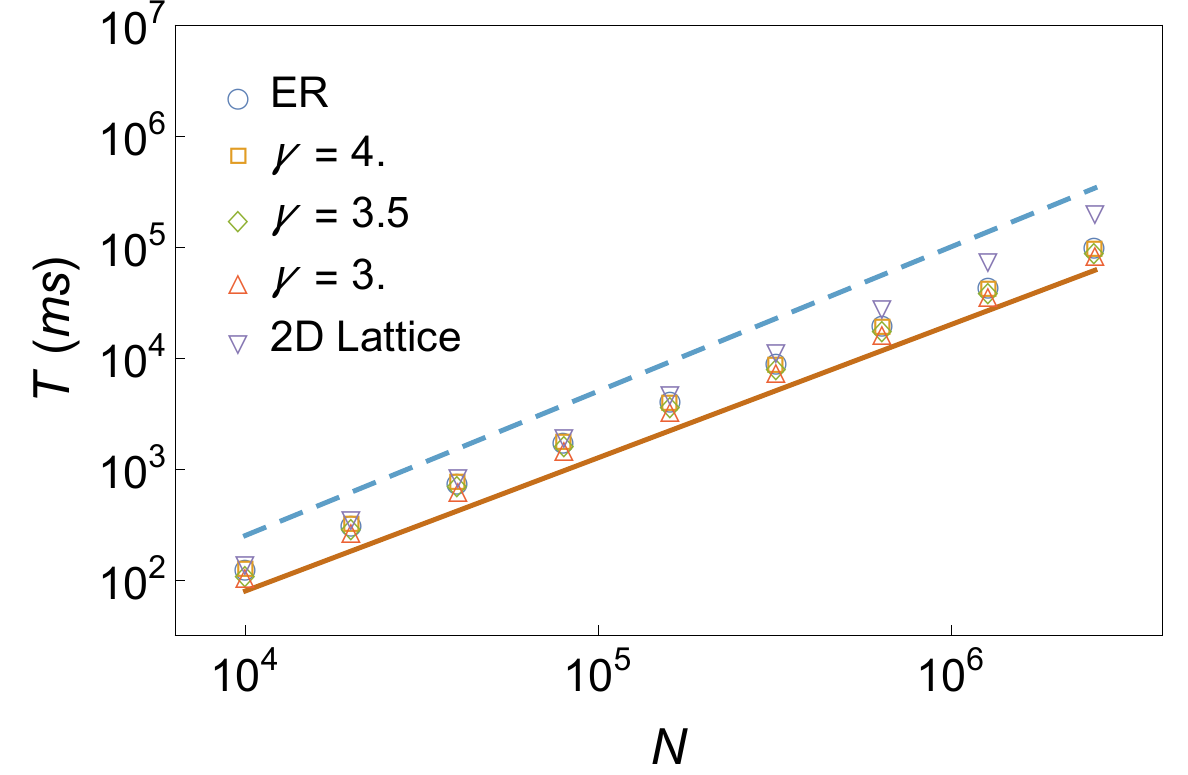}
\caption{(Color online) 
Plot of total computing time $T$ vs system size $N$ for different underlying network structures.
Each point is obtained by taking an average over 10 samples.
The solid and dashed line are guidelines with slopes  of $1.2$ and $1.3$, respectively.}
\label{Fig:ComputingTime}
\end{figure}

We measure the total computing time $T$ to keep track of the MCCs from $\KMean=4$ to 1 for the three types of  underlying networks. In \cref{Fig:ComputingTime}, we plot $T$ vs system size $N$. Each data point is obtained by averaging over 10 different samples. For ER and scale-free networks, it is likely that the degree exponent $\gamma$ does not affect time complexity for $\KMean \ge 3$, but the constant factor can vary. We find in Fig.~\ref{Fig:ComputingTime} that the time complexity depends on the system size as $\Order{N^{1.2}}$ for those complex networks.
For a two-dimensional regular lattice, we find that the time complexity is estimated as  
$T\sim \Order{N^{1.3}}$.
This numerical difference may be caused by the nature of the avalanche failures depending on the underlying network structures. 
Apart from the power-law behavior, one may expect logarithmic corrections affected by the four operations 
of the DF data structure, but this is not clearly conclusive in Fig. \ref{Fig:ComputingTime}. 

{\it Summary.}$-$ We have introduced an efficient algorithm that keeps track of the MCCs in an interdependent multiplex network. 
Our algorithm maintains the full structural information of MCCs during deletions of links, and thus it enables one to extract various 
interesting physical quantities such as the sizes of a giant MCC as well as other MCCs. 
A similar algorithm was introduced in \cite{Schneider2013}, in which, however, nodes instead of links are deleted.
In this case the algorithm can be simpler, and a multiple number of links can be simultaneously deleted. Accordingly, the computing time is 
reduced as $\Order{N\log N}$. Moreover, the algorithm was designed to trace only the largest cluster. 
In contrast, our algorithm provides other useful information on structural features of the MCCs. 
Therefore, we anticipate that our algorithm can facilitate further studies in various directions. 

Finally, we remark that the HDT algorithm utilized here can be applied to other problems such as temporal network models \cite{holme2012}
in which links can be added or deleted.

\begin{acknowledgments}
This work has been supported by  NRF Grant Nos.~2010-0015066 and 2014-069005, an SNU research grant (BK), and  the DFG within SFB 680 (SH).
\end{acknowledgments}

\appendix 
\section{ET tree and HDT algorithm} 
\label{Appendix:HDP}
\begin{figure}
\includegraphics[width=0.95\linewidth]{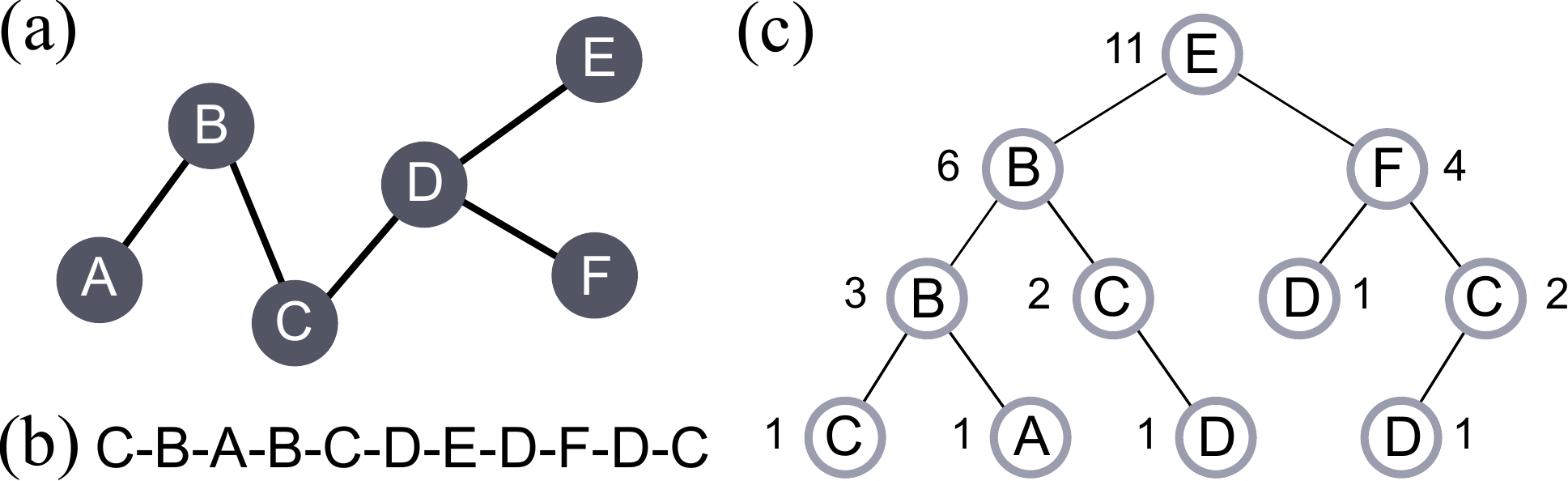}
\caption{(Color online) 
(a) The tree we want to represent. (b) An Euler tour sequence from node C of the tree. (c) The sequence stored in a balanced tree. The number next to each node of (c) is the size of the subtree from the node.}
\label{ET}
\end{figure}

The Euler tour tree is a scheme to represent a dynamical tree efficiently. For a spanning tree of size $n$, for example a tree in Fig. \ref{ET}(a), an Euler tour of the tree is a sequence of nodes recorded in a depth-first walk on the tree from an arbitrarily chosen root. It has length (the number of links) $2n-2,$ as shown in Fig. \ref{ET}(b). It starts from an arbitrary node and ends at the same node. This cycle can be represented as a sequence of $2n-1$ node indices. Each sequence is then stored in a self-balanced tree consisting of $2n-1$ nodes (Fig. \ref{ET}(c)), and its ordering is preserved. Each node of the tree carries the index of the node; thus the leftmost and rightmost nodes of each tree carry the same index. We refer to the trees built this way as Euler tour trees (Fig. \ref{ET}(c)). It is noteworthy that a node having degree $k$ in the spanning tree appears $k$ times in the Euler tour tree with one exception being the starting node, the root (node {\bf E} in Fig. \ref{ET}(c)), which appears $k+1$ times.

Note that, in  self-balanced ET trees, the hierarchical steps from the root to any terminal node are almost the same. Thus the length is $\Order{\log N}$ and the time complexity is determined accordingly.
This property is preserved even in the process of merging and splitting of the spanning trees.
There are several widely used algorithms to maintain the balance (e.g., the AVL tree or the RB tree \cite{cormen_introduction_2009}) and 
any of them can be used.

One useful piece of information would be the size of the connected component to which a given node belongs. 
For this, we augment each node of the tree to keep track of the number of its descendants. 
For example, ``11,'' the augmentation of node E in Fig. \ref{ET}(c), indicates the number of descendants. 
Whenever a node is given, identifying the root and hence finding the size of the component can be obtained in $\Order{\log N}$ steps.

After constructing such an ET tree, we run the four principal operations to the data structure $\Forest$.
Let $\Edges$ denote a set of links in the network and $\Spanning$ the set of links that constitute the forest.
Usually, $\Edges$ is the set of all links in the network, but in our algorithm it is the set of all {\em active} links.

\begin{enumerate}
\item Connected($v$,$w$, $\Forest$)
        determines whether $v$ and $w$ are in the same component.
        \begin{enumerate}
                \item Let $r_v$ and $r_w$ be the roots of the Euler trees containing $v$ and $w$, respectively.
                \item Return true if $r_v=r_w$; return false otherwise.
        \end{enumerate}
\item Size($v$, $\Forest$)
        returns the size $n$ of the component that contains $v$.
        \begin{enumerate}
                \item Find the root of the Euler tree containing $v$.
                \item The root will contain the number $s$ of its decedents, which would be $s = 2n-2$.
                        Thus return $n=s/2 + 1$.
        \end{enumerate}
\item Insert($e = (v,w)$, $\Forest$) adds a link $e$ to the DF.
        \begin{enumerate}
                \item Add $e$ to $\Edges$.
                \item If Connected($v$,$w$), do nothing. Otherwise, connect the two Euler trees adjacent to $e$.
        \end{enumerate}
\item Delete($e = (v,w)$, $\Forest$) removes a link $e$ from the DF.
        \begin{enumerate}
                \item Remove $e$ from $\Edges$. 
                \item If $e \notin \Spanning$, do nothing. If $e \in \Spanning$, remove $e$ from $\Spanning$. This will split
                        a tree into two pieces. Determine whether there exists a link $e^{\prime} \in \Edges$ that can replace $e$, i.e., connect the two again. If so, add $e^{\prime}$ to $\Spanning$.\\
        \end{enumerate}
\end{enumerate}

It is clear that Connected, Size, and Insert each require $\Order{\log N}$ steps.
In contrast, in Delete, finding $e^{\prime}$ efficiently is nontrivial. 
To achieve this, the HDT algorithm assigns an integer, called a level, for each link and maintains a spanning forest for each level.
The levels are updated during the search process of $e^{\prime}$ so that further calls of Delete can find their $e^{\prime}$ more efficiently, which sets the amortized costs of Insert and Delete to $\Order{\log^2 N}$.
For details we refer the reader to \cite{Holm2001}.

We, however, found that a brute-force searching for $e^{\prime}$ without the HDT algorithm is actually faster for our problem.
We implemented two versions of our MCC algorithm. One uses  brute-force searching and the other uses the HDT algorithm.
They exhibit the same time complexity, but the HDT algorithm has a larger prefactor in our empirical tests.
Moreover, the HDT algorithm needs additional information and consumes more memory.
Thus, we used ET trees without the HDT algorithm in the assessment of our MCC algorithm.
However, the HDT algorithm might be needed when the network size becomes much larger than the sizes used in our tests because it theoretically guarantees the $\Order{\log^2 N}$ time complexity while we cannot provide a precise time complexity for the brute-force searching.

\section{Successive removal of link $e$ from $\Added_A$}
\label{Appendix:Removal}
\begin{enumerate}
\item If $e\in\Inactive_A$, remove $e$ from $\Inactive_A$.

\item Otherwise, perform Delete($e=(u,v),\Forest_A$).
        This might split an MCC into two. 
        Check whether this occurs using Connect($u,v,\Forest_A$).
\begin{enumerate}
\item If Delete does not split any connected component, nothing more needs to be done.
\item If it does, some links in $\Active_B$ have to be changed to inactive, as one can see from \cref{Fig:diagram}.
\begin{enumerate}
                        \item All links in $\Active_B$ that connect the two split components will become inactive.
                                To find them, one can scan each node in the smaller component exhaustively and determine whether any of its
                                        outgoing links connect it to a node belonging to the larger component.
                        \item For each link $e$ of these, perform Delete($e, \Forest_B$) 
                                and add it to $\Inactive_B$.
                        \item Of course, this, in turn, can trigger some links of $\Active_A$ into becoming 
                        inactive, and again we perform Delete for each.
                        \item This recursive process has to be performed until no more inactive links are generated.
                \end{enumerate}
        \end{enumerate}

\end{enumerate}
\bibliography{references0} 

\end{document}